\documentstyle[11pt,epsf,vsopsymp]{article}
\def\tj{$\theta_{\rm{jet}}$}
\def\bj{$\beta_{\rm{jet}}$}
\begin{document}


\markboth{\hfill Sudou et al.}{The Counter Jet in NGC 6251 \hfill}


\centerline{\LARGE\bf 
   Detection of the VLBI-scale Counter Jet 
}\bigskip

\centerline{\LARGE\bf 
   in NGC 6251
}\bigskip

\centerline{\sc 
H.~Sudou$^1$, Y.~Taniguchi$^1$, Y.~Ohyama$^2$, S.~Kameno$^2$,
}
\centerline{\sc 
S.~Sawada-Satoh$^2$, M.~Inoue$^2$, O.~Kaburaki$^1$, \& T.~Sasao$^2$,
}\medskip

\centerline{\it
$^1$ Astronomical Institute, Tohoku University, Aoba, Sendai  980-8578, Japan 
}
\centerline{\it
$^2$ NAO, Osawa 2-21-1, Mitaka, Tokyo 181-8588, Japan
}

\begin{abstract} \noindent 
Mapping the central 5 pc region of the nearby radio
galaxy NGC 6251  
with a 0.2 pc resolution using VLBI at two radio frequencies, 5 GHz
and 15 GHz, we have found the sub-pc-scale counter jet
for the first time in this radio galaxy. This discovery allows us to
investigate the jet acceleration based on the relativistic beaming
model (Ghisellini et al. 1993).
\end{abstract}


\sources{NGC 6251}

\section{Introduction}
The genesis of powerful radio jets from active galactic nuclei is one of the
long standing problems in astrophysics
(Bridle \& Perley 1984).
Although global morphological properties give us very
important 
information, very inner regions in the radio jets also provide
hints to understanding the genesis of radio jets.
In order to investigate
radio galaxies at the sub-milli arcsecond angular resolution, 
we have performed new high-resolution VLBI observations of NGC 6251
using HALCA (Hirabayashi et al. 1998).   
We use a distance to NGC 6251, 94.4 Mpc 
(for a Hubble constant  $H_0$ = 75 km s$^{-1}$ Mpc$^{-1}$).
Note that 1 mas (milli arcsecond) corresponds to 0.48 pc at this
distance.

\section{Observations}

NGC 6251 was observed at 5 GHz using VSOP on 30 April 1998 
 and at 15 GHz using VLBA on 2 June 1998.
Details of the observations are summarized in Table
1. 
In order to perform
beam-size-matched comparison between 5 GHz and 15 GHz, we restored the
two images with a same spatial resolution of 0.50 $\times$ 0.50 mas.
\begin{table}[t]
\caption{Observations. } 
 \begin{footnotesize}
\begin{tabular}{cccccc}
\hline \hline
 Date&
 $\nu$&
 Stations&
 Peak Int.&
 RMS noise&
 DR$^1$\\
 &
 (GHz)&
 &
 {\small (Jy/beam)}&
 {\small (mJy/beam)}&
 \\
 \hline
 1998 Apr. 30 & 5  &  VLBA, HALCA, EB$^2$ &  0.13 & 0.50 & 260 \\
 1998 Jun. 02 & 15 & VLBA &                  0.34 & 0.25 & 1400 \\
\hline \hline
\end{tabular}
  \end{footnotesize}
 \begin{footnotesize}
 $^1$DR : Dynamic range; 
 $^2$EB : Effelsberg 100-m telescope
  \end{footnotesize}
\end{table}

\section{Results}

Our final maps at 5 GHz and 15 GHz are shown in Figure 1a and 1b,
respectively. 
Although
the secondary peak is seen at both frequencies, 
its angular
distance from the 15 GHz brightest peak is 
larger by 0.3 mas (0.14 pc) than that at 5 GHz.
This difference may be attributed to the different opacity 
toward the central engine at 5 GHz and 15 GHz.

\begin{figure}
 \begin{center}
\epsfile{file=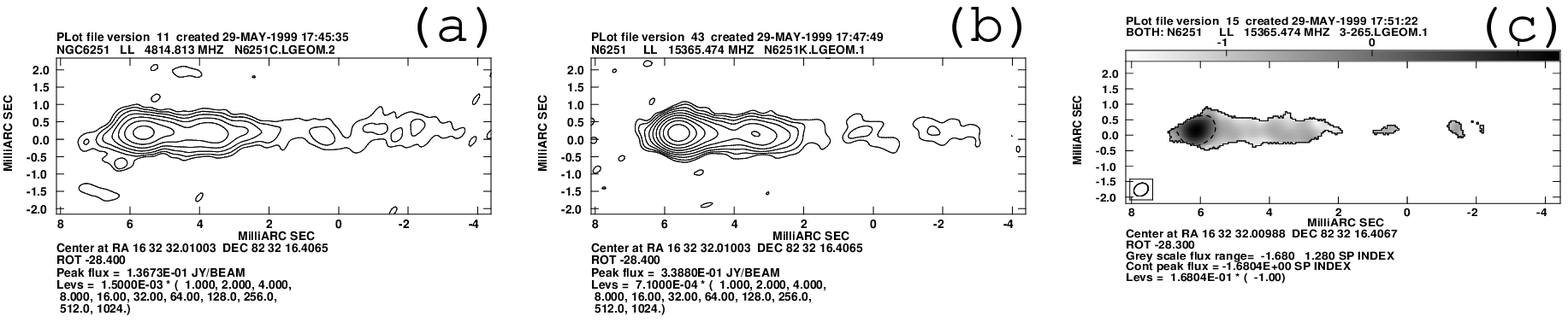,scale=0.7}
\caption{The images of NGC 6251 at 5 GHz (a) and at 15 GHz
  (b).
The 
images are rotated clockwise on the sky by $28^\circ$. The spectral
  index map is also shown (c). Here the spectral index $\alpha$ is
  defined as $S_\nu \propto  
\nu^{\alpha}$. }
 \label{Fig1}
 \end{center}
\end{figure}

Here we apply two new methods for the registration using correlation
functions between intensity profiles and between wiggling patterns of
the radio jet at the two frequencies. 
The correlation function of intensity profiles exhibits the maximum
correlation 
coefficient $\simeq 0.98$ at an offset of 
0.30 $\pm$ 0.11 mas, indicating that the peak at 5 GHz is shifted 
to the jet direction from that at 15 GHz.
Analysis of wiggling patterns gives an offset of 
 0.60
$\pm$ 0.36 mas  with a correlation coefficient
of $ \simeq$0.66. Since these two offsets were obtained using 
the two independent methods, we adopt a weighted average of the 
two offsets. Thus the best offset of 0.42 $\pm$ 0.14 mas is obtained. 
Using this offset, we register the two images and obtain a
spectral index image of the radio jet as shown in Figure 1c.
It can be seen optically thin component in the opposite side of the jet. We
conclude that this is the real counter jet.

\section{Discussions}

We assume for simplicity that the core is surrounded by a plasma sphere 
with a radius of $a$ and the radio emission from the inner part of
the jets (i.e., both the jet and the counter jet) suffer from 
the free-free absorption. The
  approaching jet escapes from the plasma sphere at a projected  
distance of $x = a \times \sin \theta_{\rm jet} $
  and thus  the 
spectral index becomes to be intrinsic here.
The counter jet suffers the effect of
free-free absorption until $x = -a $.
It is also noted that
the path length is longest at $x =
  - a \times \sin \theta_{\rm jet} $. As shown in Figure 2a, we define
  the  following projected distances;
$X_{\rm jet} = X_{\rm peak} =  a \times \sin \theta_{\rm jet}$
and $X_{\rm cjet} = a$. 
Then we are able to estimate 
$ \theta_{\rm jet} = \sin^{-1}
  \left({\frac {X_{\rm jet}}{X_{\rm cjet}}} \right)
$. We adopt the
15 GHz brightest peak as the core, because the optical depth at 15 GHz
is enough to be small in the case of free-free absorption.
In Figure 2b, we show the observed spectral index variation
along the radio jet. We estimate $X_{\rm peak} \approx$ 0.24 mas,
$X_{\rm jet} \approx$ 0.41 mas, and $X_{\rm cjet} \approx$ 0.80 mas.
Then we obtain $\theta_{\rm jet} \simeq 31^\circ$.

\begin{figure}
 \begin{center}
\epsfile{file=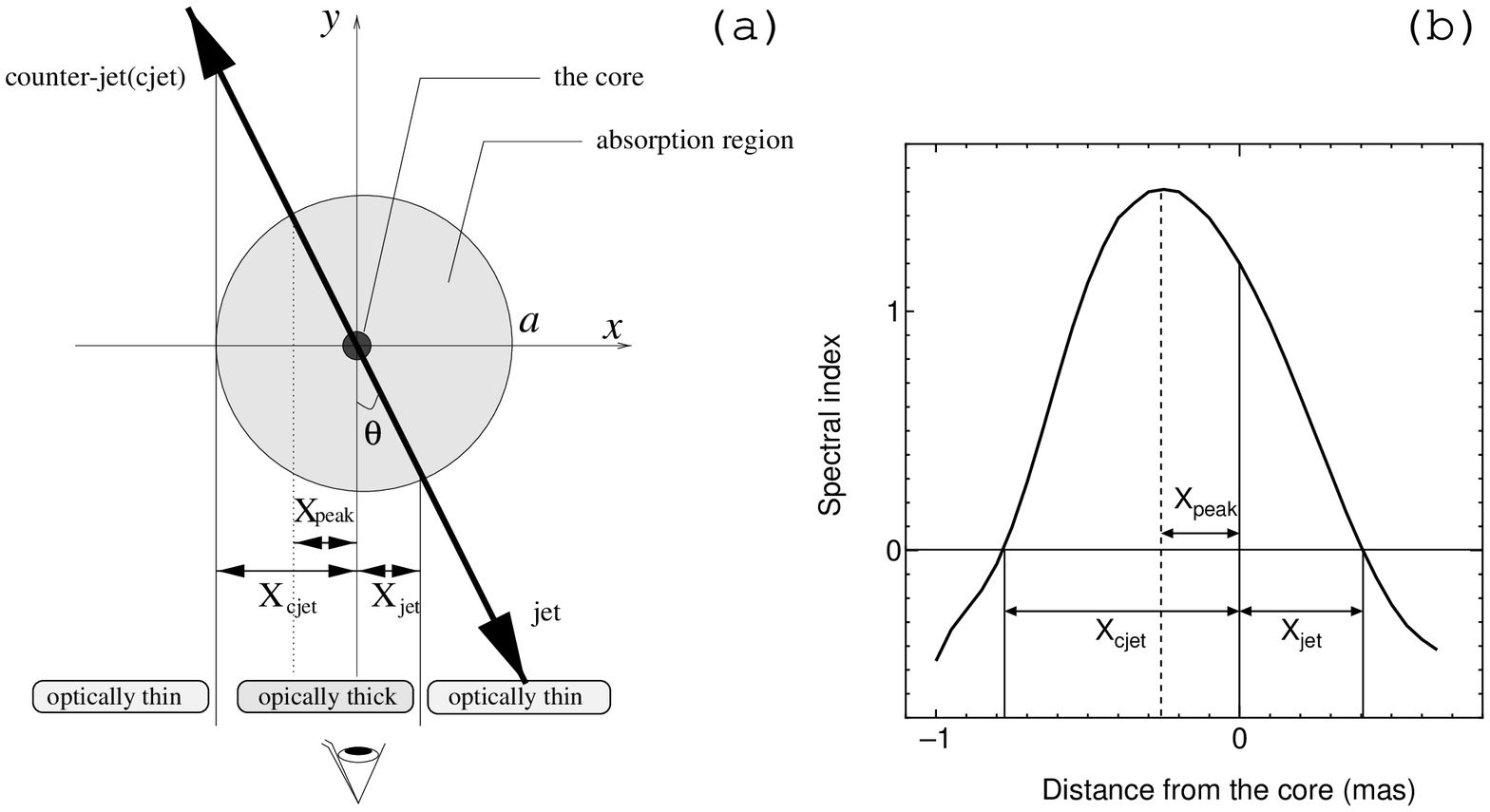,scale=0.52} 
\caption{A schematic illustration of the spherical-absorption
  model (a), and spectral index variation along the radio jet of NGC 6251 (b).}
 \label{Fig2}
  \end{center}
\end{figure}

\begin{figure}
 \begin{center}
\epsfile{file=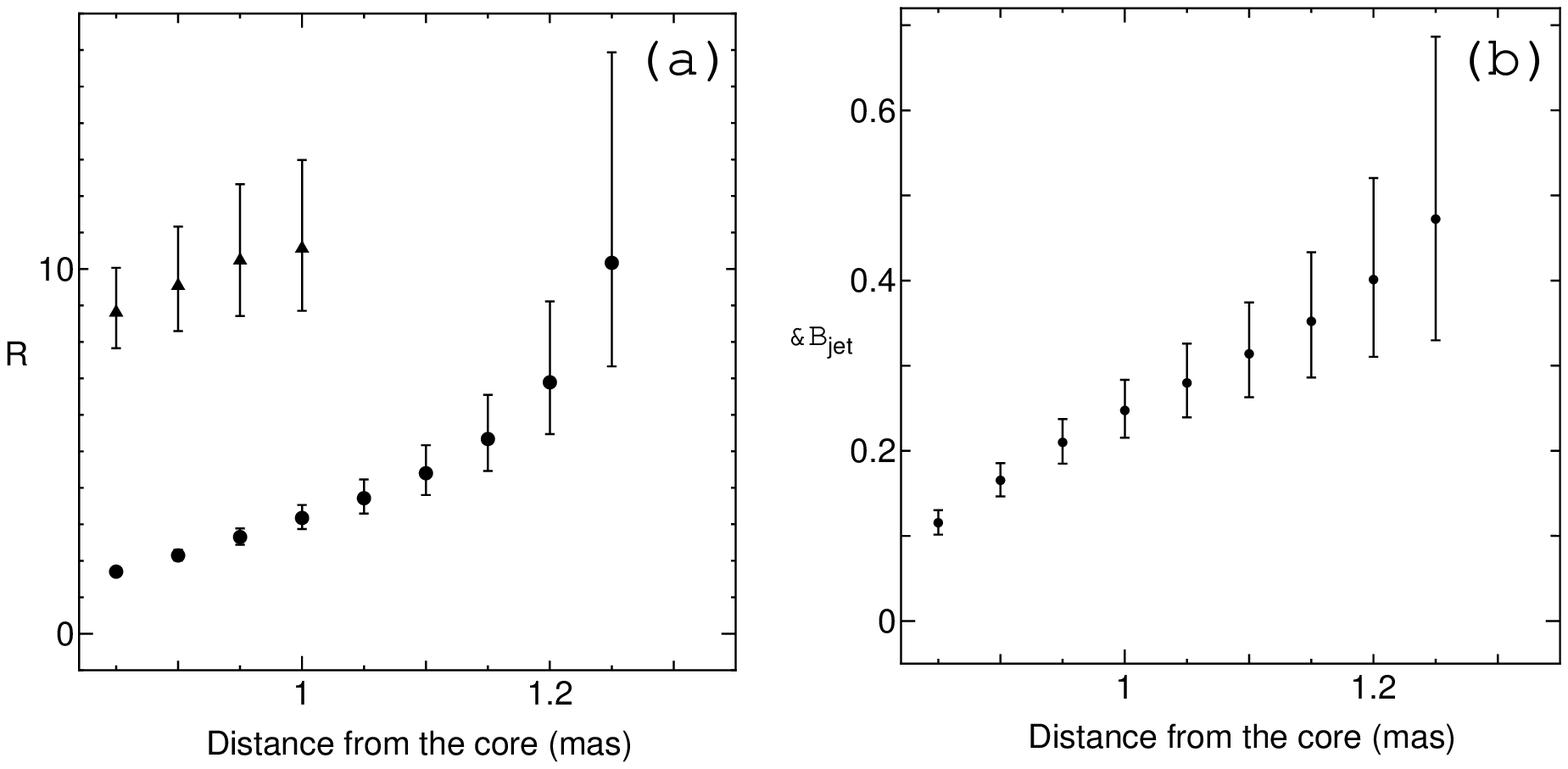,scale=0.46}
\caption{The jet/counter jet flux density ratio $R$ as a function of 
the projected distance from the brightest peak at 15 GHz for the 5GHz data 
({\it filled triangle}) and for the 15 GHz data 
({\it filled circles}) (a), and jet velocity \bj ~ as a function of
 projected 
distance from the brightest peak at 15 GHz (b).}
 \label{Fig2}
  \end{center}
\end{figure}

Adopting the so-called Doppler beaming model, 
the jet to the counter jet intensity ratio $R$ can be written as a
function of projected distance $x$, 
$R(x) = \left[\frac {1 + \beta_{\rm{jet}}(x) \cos \theta_{\rm{jet}}(x)}
   {1 - \beta_{\rm{jet}}(x) \cos \theta_{\rm{jet}}(x)}\right]^{2 -
   \alpha(x)}$ , 
where \bj($x$) = $v_{\rm jet}(x) / c$.
In Figure 3a,
we measured $R$ as a function of $x$. 
It is shown that $R$ is estimated to be systematically 
larger at 5 GHz than those at 15 GHz. 
This is
probably due to stronger absorption at 5 GHz
than at 15 GHz.
It is shown that $R$ at 15 GHz increases from 1.7 at 0.85 mas  (0.4 pc)
to 10 at 
1.25 mas (0.6 pc) with  
projected distance from the core.
If this increase is just caused by the Doppler beaming, it is suggested
that \bj ~ increase with  distance, 
because 
it cannot be seen that \tj ~ varies significantly.
Figure 3b shows that the jet is accelerated from \bj ~
$\approx 0.1$ at 
0.4 pc to \bj ~ $\approx 0.5$ at 0.6 pc.
This provides the first direct evidence for the
acceleration at the sub-pc-scale radio jet.

\acknowledgements

We gratefully acknowledge many staff members of VLBA and VSOP who operated 
observations without whom this study would not have been
possible.

\end{document}